\documentclass[11pt]{article}
\usepackage{amsmath,amssymb}
\usepackage{graphicx} 
\usepackage{bbm}
\usepackage{longtable}
\usepackage[small,loose]{subfigure}
\usepackage[small]{caption2} 
\usepackage{fleqn} 
\usepackage{mathrsfs}   
\usepackage{pst-all}
\usepackage{scrtime}
\usepackage{pifont}
\usepackage{cite}

\advance \headheight by 3.0truept       
\setcaptionwidth{0.75\textwidth}

\newcommand{\CenterEps}[2][1]{\ensuremath{\vcenter{\hbox{\includegraphics[scale=#1]{#2.eps}}}}} 

\newcommand{\E}[1]{\ensuremath{\mathrm{E}_{#1}}} 

\newcommand{\SO}[1]{\ensuremath{\mathrm{SO}(#1)}}
\newcommand{\SU}[1]{\ensuremath{\mathrm{SU}(#1)}}
\newcommand{\U}[1]{\ensuremath{\mathrm{U}(#1)}}
\newcommand{\Z}[1]{\ensuremath{\mathbbm{Z}_{#1}}} 

\newcommand{\bs}[1]{\boldsymbol{#1}}

\allowdisplaybreaks

\begin{document}

\title{
\begin{flushright}
\normalsize{CERN-PH-TH/2008-157}\\
\normalsize{TUM-HEP 692/08}
\end{flushright}
\vskip 2cm
{\bf\huge Heterotic mini-landscape (II): completing the search for MSSM vacua in
a $\boldsymbol{\Z6}$ orbifold}\\[0.8cm]}

\author{{\bf\normalsize
Oleg~Lebedev$^{1}$\!,
Hans Peter~Nilles$^2$\!,
Sa\'ul~Ramos-S\'anchez$^2$\!,}\\{\bf\normalsize
Michael~Ratz$^3$\!\!,
Patrick~K.S.~Vaudrevange$^2$\!} \\[1cm]
{\it\normalsize
${}^1$ CERN, Theory Division, CH-1211 Geneva 23, Switzerland}\\[0.05cm]
{\it\normalsize
${}^2$ Physikalisches Institut der Universit\"at Bonn,}\\[0.05cm]
{\it\normalsize Nussallee 12, 53115 Bonn,
Germany}\\[0.1cm]
{\it\normalsize
${}^3$ Physik Department T30, Technische Universit\"at M\"unchen,}\\[0.05cm]
{\it\normalsize James-Franck-Stra{\ss}e, 85748 Garching,
Germany} 
}
\date{}
\maketitle 
\thispagestyle{empty} 
\begin{abstract}
{ We complete our search for MSSM vacua in the \Z6-II heterotic orbifold by
including models with 3 Wilson lines. We estimate the total number of inequivalent models
in this orbifold to be $10^7$. Out of these, we find almost 300 models with the exact MSSM
spectrum, gauge coupling unification and a heavy top quark. Models with  these features 
 originate predominantly  from  local GUTs. The scale of gaugino condensation in the hidden sector
is correlated with properties of the observable sector such that soft masses in the TeV range are
preferred. 
}
\end{abstract}
\clearpage

\section{Introduction}

Construction of the (supersymmetric) standard model vacua  has been one of the
top priorities in string theory. Although there is a vast landscape of string
theory vacua \cite{Lerche:1986cx,Susskind:2003kw},  realistic models are
extremely rare. For example, in Gepner models the probability of finding a model
with the massless spectrum of  the minimal supersymmetric standard model (MSSM) 
plus vector--like exotics  is of order $10^{-14}$ \cite{Dijkstra:2004cc} while
this probability in \Z6  intersecting brane models is $10^{-16}$
\cite{Gmeiner:2007zz} (for models with chiral exotics it is $10^{-9}$
\cite{Gmeiner:2005vz}). The heterotic string
\cite{Gross:1984dd,Gross:1985fr,Dixon:1985jw,Dixon:1986jc,Ibanez:1986tp,Ibanez:1987xa,Ibanez:1987sn,
Casas:1988hb,Casas:1987us}, on the other hand, provides a more fertile ground
for realistic constructions due to its built--in grand unification structures.
In particular, there are fertile regions in the \Z6-II heterotic orbifold with 
2 Wilson lines where the probability of finding a model with the \emph{exact}
MSSM spectrum is somewhat below 1\,\%  \cite{Lebedev:2006kn} and about 100 such
models have been  identified (see also
\cite{Bouchard:2005ag,Braun:2005nv,Blumenhagen:2006ux,Kim:2006hv,Kim:2007mt}).
In this paper, we complete our search \cite{Lebedev:2006kn} within the \Z6-II
heterotic orbifold by including models with 3 Wilson lines, which is the maximal
possible number of Wilson lines in the \Z6-II orbifold. This allows us to
estimate the total number of inequivalent models in this  orbifold and construct
further examples of MSSMs. Unlike in the presence of 2 Wilson lines, all 3
matter generations are fundamentally different in this case. 
 
In the first part of our analysis, we consider the gauge shifts associated with
\SO{10} or \E6 local grand unified theories (GUTs). This is the strategy we have
pursued in our previous paper \cite{Lebedev:2006kn}. Inspired by an orbifold GUT
interpretation of heterotic models
\cite{Kobayashi:2004ud,Forste:2004ie,Kobayashi:2004ya,Buchmuller:2004hv}, 
\emph{local GUTs} 
\cite{Buchmuller:2004hv,Buchmuller:2005jr,Buchmuller:2006ik,Buchmuller:2005sh,Buchmuller:2007qf}
are specific to certain points in the compact space such that twisted states
localized at these points  form complete representations of the corresponding
GUT group. On the other hand, the 4D gauge symmetry is that of the SM and the
bulk states such as gauge bosons (and Higgs doublets) only form representations
of the latter.  This provides a heuristic explanation of the apparent GUT 
structure of the SM matter multiplets without having a 4D GUT.   The MSSM search
strategy based on local GUT gauge shifts has been very successful and led to
identification of about 100 models with the exact MSSM spectrum
\cite{Lebedev:2006kn}. All these models involve 2 Wilson lines and share  the 
common feature that 2 matter generations are very similar, while the third one
is fundamentally different. They all have the top quark  Yukawa coupling of
order one and are consistent with gauge coupling unification. In our current
work, we extend these results to models with 3 Wilson lines and construct
further ${\cal O}$(100) models  with the MSSM spectrum. In this case, all 3
matter generations are  different which leads to distinct (but not necessarily
``healthier'') phenomenology.  

In the second part of our analysis, we relax the requirement of having \SO{10}
or \E6 local GUTs and construct MSSMs based on arbitrary gauge shifts. This is
interesting as it allows us to determine how likely is a given model with the
MSSM  spectrum to have originated from a local GUT. 
Finally, we provide a
representative example of a 3 WL model with the exact MSSM spectrum.

\section{Constructing MSSMs}

In our previous mini-landscape study \cite{Lebedev:2006kn}, we have analyzed
models with up to 2 Wilson lines and  local \SO{10} and \E6 structures.  Let us
briefly review the key ingredients of this construction (for more details, see
\cite{Buchmuller:2006ik,Nilles:2008gq}). An orbifold model is
defined by the orbifold twist, the torus lattice and the gauge embedding of the
orbifold action, i.e.\ the gauge shift $V$ and the Wilson lines $W_n$.  The
\Z6-II orbifold allows us to switch on one Wilson line of degree 3 ($W_3$) and
up to two of degree 2 ($W_2$ and $W_2'$). A given $V$ corresponds to the 
\SO{10} or \E6 local GUT if the left-moving momenta $p$ satisfying
\begin{equation} 
 p \cdot V~=~0 \mod 1, ~~p^2~=~2
\end{equation}
are roots of \SO{10} or \E6 (up to extra gauge factors). In addition, this $V$
must allow for massless  $\boldsymbol{16}$--plets of \SO{10} at the fixed points with \SO{10}
symmetry or  $\boldsymbol{27}$--plets of \E6 at the fixed points with \E6
symmetry. Since massless states from $T_1$ are automatically invariant under the
orbifold action, they all survive in 4D and appear as complete GUT  multiplets.
In the case of \SO{10}, that gives one complete SM generation, while in the
case  of \E6 it is necessary to decouple part of the $\boldsymbol{27}$--plet
since  $\boldsymbol{27}=\boldsymbol{16}+\boldsymbol{10}+\boldsymbol{1}$ under
\SO{10}. Then, choosing appropriate Wilson lines $W_n$, one obtains the SM gauge
group in 4D. Furthermore, in order to have the correct hypercharge
normalization, one  requires the embedding $G_\mathrm{SM} \subset \SU5$. 

In the \Z6-II orbifold, there are 2 gauge shifts leading to a local \SO{10}
GUT,
\begin{eqnarray}
V^{ {\rm SO(10)},1}& = &
\left(\tfrac{1}{3},\,\tfrac{1}{2},\,\tfrac{1}{2},\,0,\,0,\,0,\,0,\,0\right)~\left(\tfrac{1}{3},\,0,\,0,\,0,\,0,\,0,\,0,\,0\right) \;,
\nonumber \\ 
V^{ {\rm SO(10)},2 }& = &
\left(\tfrac{1}{3},\,\tfrac{1}{3},\,\tfrac{1}{3},\,0,\,0,\,0,\,0,\,0\right)~\left(\tfrac{1}{6},\,\tfrac{1}{6},\,0,\,0,\,0,\,0,\,0,\,0\right) \;,
\label{eq:so10shifts}
\end{eqnarray}
and 2 shifts leading to a local \E6 GUT,
\begin{eqnarray}
V^{\E6 , 1}= &
\left(\tfrac{1}{2},\,\tfrac{1}{3},\,\tfrac{1}{6},\,0,\,0,\,0,\,0,\,0\right)&\left(0,\,0,\,0,\,0,\,0,\,0,\,0,\,0\right)\;,
\nonumber \\ 
V^{ \E6 ,2}= &
\left(\tfrac{2}{3},\,\tfrac{1}{3},\,\tfrac{1}{3},\,0,\,0,\,0,\,0,\,0\right)&\left(\tfrac{1}{6},\,\tfrac{1}{6},\,0,\,0,\,0,\,0,\,0,\,0\right).\label{eq:e6shifts}
\end{eqnarray}
Having fixed these shifts, one scans over possible Wilson lines to get the SM
gauge group. To identify MSSM candidates, we have taken the following steps:

\begin{dingautolist}{"0C0}
  \item Generate Wilson lines $W_3$ and $W_2$
  \item Identify ``inequivalent'' models
  \item Select models with $G_\mathrm{SM} \subset \SU5 \subset \SO{10} $ or \E6
  \item Select models with three net ({\bf 3,2})
  \item Select models with non--anomalous $\U1_{Y} \subset \SU5$
  \item Select models with net 3 SM families + Higgses + vector--like exotics
  \item Select models with a heavy top 
  \item Select models in which the exotics decouple
\end{dingautolist}

The result was that out of $3\times 10^4$ inequivalent models about 100 models
satisfied our MSSM--requirements. Thus, close to 1\% of all models were 
acceptable. These models have 2 identical matter generations from two localized
$\boldsymbol{16}$- or $\boldsymbol{27}$--plets, which is due to the presence of
one Wilson line of order three ($W_3$) and one Wilson line of order two
($W_2$).  

In this work, we extend our previous analysis by allowing for 3 Wilson lines,
which is the maximal possible number of Wilson lines in the \Z6-II orbifold.  
An immediate consequence of this  is that all three matter generations  obtained
in this case would have a different composition. Also, since (due to
combinatorics) most models in the \Z6-II orbifold have 3 Wilson lines, this
allows us to estimate  the total number of all possible models and the
probability of finding the MSSM by a ``blind scan''. Furthermore, we relax the
requirement of the hypercharge embedding into a local \SO{10} or \E6 GUT, while
still having the correct GUT hypercharge normalization. Finally, we drop the
requirement of  having a local \SO{10} or \E6 GUT. Besides constructing new
models, all this helps us understand whether (and how) the ``intelligent''
search strategy  based on local GUTs is more efficient than a ``blind scan''.
Also, given a model with the exact MSSM spectrum, gauge coupling unification and
a heavy top quark, we can determine how likely it is to have come from a local
GUT.

\subsection{3 WL models with local GUTs}

We start by studying the models with local GUT shifts of \cite{Lebedev:2006kn}.
Our results are presented in Tab.~\ref{tab:Summary_3WL}.  Note the difference 
in step \ding{"0C2} compared to that in  the 2 Wilson line case: now we do not
require  the hypercharge embedding  in \SU5$~\subset~$\SO{10}  at this step,
whereas  at step \ding{"0C4} we require   $\U1_{Y}\subset \SU5 $ with \SU5 not
necessarily being inside \SO{10} (or \E6). This allows us to retain more models while
keeping the standard GUT hypercharge normalization.

Compared to the 2 WL case, the total number of inequivalent models has grown
from $3\times 10^4$ to $10^6$. In the end, however, we retain only about 100
new models. Thus the efficiency is much lower than that in the 2 WL case. It is
interesting that most of the models at step \ding{"0C7} come form the \E6 local
GUT with the gauge shift $V^{\E6,1}$. The fact that \E6 models contribute much
more in the 3 WL  than 2 WL case is understood by symmetry breaking: it is 
easier to get to the SM gauge group from \E6 using 3 Wilson lines.

\begin{table*}[h!]
\begin{center}
\begin{tabular}{l||r|r||r|r}
 criterion & $V^{\SO{10},1}$ & $V^{\SO{10},2}$ & $V^{\E6,1}$ & $V^{\E6,2}$\\
\hline
& & & & \\
\ding{"0C1} ineq. models with 3 WL
  & $942,469$ & $246,779$ & $8,815$ & $37,407$ \\[0.2cm]
\ding{"0C2} SU(3)$\times$SU(2) gauge group
  & $373,412$ & $89,910$  & $2,321$ & $13,857$\\[0.2cm]
\ding{"0C3} 3 net $(\boldsymbol{3},\boldsymbol{2})$
  & $5,853$   & $2,535$   & $352$ & $745$\\[0.2cm]
\ding{"0C4} non--anomalous $\U1_{Y}\subset \SU5 $
  & $2,620$   & $1,294$   & $314$ & $420$\\[0.2cm]
\ding{"0C5} spectrum $=$ 3 generations $+$ vectorlike
  & $45$ & $19$ & $123$ & $0$\\[0.2cm]
\ding{"0C6}  heavy top
  & $44$ & $1$ & $123$ & $0$\\[0.2cm]
\ding{"0C7}  exotics decouple at order 8
  & $20$ & $1$ & $60$ & $0$ \\
\hline
\end{tabular}
\end{center}
\caption{Statistics of $\Z6-$II orbifold models based on the shifts
 $V^{\SO{10},1},V^{\SO{10},2},V^{\E6,1},V^{\E6,2}$ with three 
 non--trivial Wilson lines.}  
 \label{tab:Summary_3WL}
\end{table*}

A comment is in order. Due to our computing limitations,  we define two models
to be ``equivalent'' if they have identical non--Abelian massless spectra and
the same number of non--Abelian massless singlets. This does not take into
account the possibility that the singlets can have different U(1) charges, nor
that fields with identical gauge quantum numbers can differ in their
localization, etc. These differences can sometimes be important, for example,
for the decoupling of exotics since the relevant mass terms can be allowed in
one case and not the other. As a result, we underestimate the number of
inequivalent models. The resulting uncertainty in our numbers is found
empirically to be within a factor of 2.

\subsection{A statistical analysis of general 3 WL models}

Now we turn to the discussion of general 3 WL models, i.e.\ we no longer demand
that there is a local \SO{10} or \E6.
The number of models with 3 Wilson lines is very large and, unlike in the case
of 2 Wilson lines, constructing all of them (in the sense of calculating the
spectrum) is an extremely time--consuming task. The reason is that the known
ways of constructing all inequivalent models lead to huge redundancies because
different shifts and Wilson lines can be related by elements of the Weyl group 
(cf.\ the discussion in \cite{Giedt:2000bi}).
Thus, it becomes impossible to check how many of them are equivalent. Instead of
the complete classification of models we use a statistical approach (for related
discussion see~\cite{Dienes:2006ut,Dienes:2006ca,Dienes:2007ms}). To understand
the basic idea,  consider a simple example. Suppose we have a set of $M$ models
out of which $N$ are inequivalent ($M,N \gg 1$). Assume also that each
inequivalent model is represented $M/N$ times in the set $M$, which corresponds 
a ``flat distribution''. The probability that 2 randomly chosen models are
equivalent is $1/N$. Take now a larger random selection  of models  $n$, $1 \ll
n \ll N$. The probability that there are equivalent models in this set is
\begin{equation}
 p(n,N)~\simeq~\binom{n}{2} \frac{1}{N}~\simeq~\frac{n^2}{2 N} \;.
\end{equation}
For $n=\sqrt{N}$, this probability is 1/2. Thus, in a sample of $\sqrt{N}$ out
of a total of $N$ models, there is order 1 probability that at least 1 model is
redundant. This observation  allows us to estimate the number of inequivalent
models by studying a sample of order $\sqrt{N}$ models.

As the first step, we use the following simple algorithm. Start with a random
model. Then  generate another  model and compare it to the first one. If they
are  equivalent, stop the procedure. Otherwise, generate another model and
compare it to the previous ones, and so on. The probability  that this procedure
terminates at a sample of size $n$ is
\begin{equation}
 P(n,N)~=~\frac{n}{N}\, \prod_{k=1}^{n-1} \left( 1-\frac{k}{N} \right) \;.
\end{equation}
The maximum of this function is at $n=\sqrt{N}$. Thus, if we produce a number of
sets of models with different $n$ and plot how common a particular $n$ is, the
maximum of this distribution should give $n=\sqrt{N}$.

An important assumption in this analysis is that all  inequivalent models are
equally likely to be generated. In practice, this is not the case and some
models appear more often than the others. This, in particular, has to do with 
the specifics of the  model--generating routine. To take this factor into
account, we introduce a fudge parameter $t$ defined by $n\, t  \simeq \sqrt{N}$,
where $n$ is the predominant size of the sample (as defined above) and $N$ is
the true number of inequivalent models. The parameter $t$ can be determined
``experimentally'' when both $n$ and $N$ are known, for example, from the 2
Wilson line case distributions or subsets of 3 Wilson line case distributions.
We find a rather stable value $t \simeq 2$  independently of the sample
considered and adopt this value for the rest of our analysis.

Using these methods, we consider models with all possible gauge shifts, and 2
and 3 non--trivial Wilson lines. 
We find that there are about $10^7$ inequivalent models.
Out of these, we have constructed explicitly all possible models with 2 WLs and
a sample of $5\times 10^6$  models with 3 WLs.  
This resulted in 267 MSSMs.\footnote{Here we only obtain 74 MSSMs with 3 WL
which is fewer than the number in Tab.~\ref{tab:Summary_3WL}. This is because
our sample of $5\times 10^6$ models does not contain all models with local
\SO{10} and \E6. } Most of them originate from  \E6, \SO{10} and \SU5 local
GUTs as shown in Tab.~\ref{tab:LocalGUTStructurePromisingModels}. Note that
models with \SU5 local structure do  not have a complete localized family,
rather only part of it. The additional states come from other sectors of the
model.  The conclusion is that any model with  the exact MSSM spectrum, gauge
coupling unification and a heavy top quark is likely to have come from some
local GUT.

\begin{table*}[h!]
\centerline{
\begin{tabular}{c|c|r|r}
 local GUT & ``family'' &  2 WL & 3  WL \\
\hline
&&&\\
\E6          & $\bs{27}$ & $14$  & $53$ \\[2mm]
\SO{10}      & $\bs{16}$ & $87$  &  $7$ \\[2mm]
\SU6         & $\bs{15}$+$\bs{\bar 6}$ &  $2$  &  $4$ \\[2mm]
\SU5         & $\bs{10}$ & $51$  & $10$ \\[2mm]
rest &                 & $39$  &  $0$ \\
\hline
total &                 & $193$ & $74$ \\
\hline
\end{tabular}
}
\caption{Local GUT structure  of the  MSSM candidates. These gauge groups
appear at some fixed point(s) in the $T_1$ twisted sector. The \SU5 local GUT
does not  produce a complete family, so additional ``non-GUT'' states are 
required.}
\label{tab:LocalGUTStructurePromisingModels}
\end{table*}

\begin{figure}[!h!]
\subfigure[]{
\CenterEps[0.75]{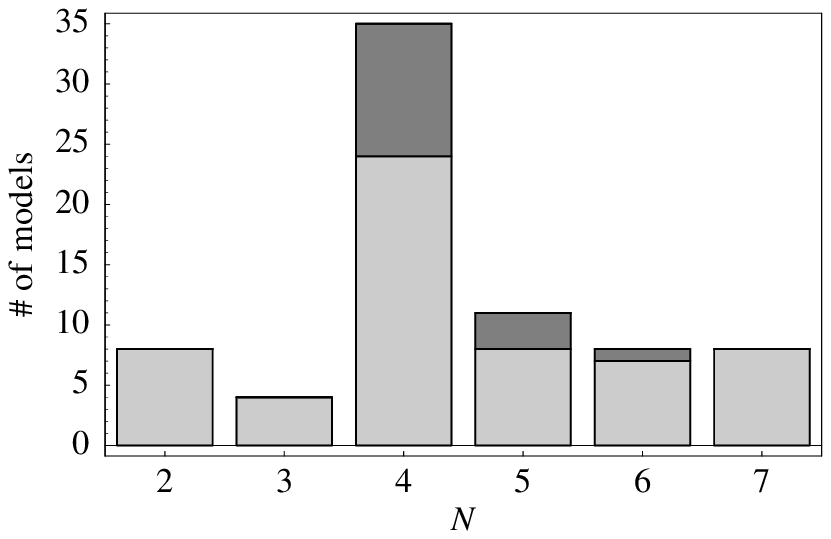}
}
\qquad
\subfigure[]{
\CenterEps[0.75]{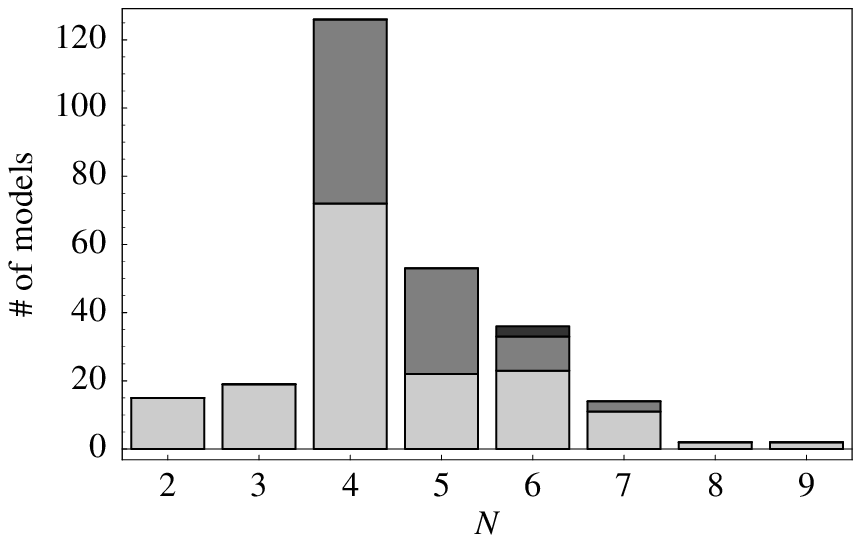}
}
\caption{ Number of MSSM candidates vs.\ largest gauge group in the
hidden sector. \SU{N}/\SO{2N}/\E{N} are given by light/dark/darker bins. (a):
models with 3 WL, (b): models with 2 and 3 WL.}
\label{fig:ComparingHiddenSector3WLVs2-3WL}
\end{figure}

An important characteristic of MSSM candidates is the size of the hidden sector
which determines the scale of gaugino condensation $\Lambda$  and consequently
the scale of soft SUSY breaking masses $m_{3/2}$
\cite{Nilles:1982ik,Ferrara:1982qs,Derendinger:1985kk,Dine:1985rz},  
\begin{equation}
 m_{3/2}~\sim~\frac{\Lambda^3}{M_\mathrm{P}^2}\;,
\end{equation}
where $M_\mathrm{P}$ denotes the Planck scale. If the largest hidden sector
gauge factor  is too big, e.g. \E6  or \E8, the gaugino condensation scale is
too high and supersymmetry is irrelevant to low energy physics. If it is too
low, the model is ruled out by experiment.  It is intriguing that, in the 2 WL
case, most of the MSSM-candidates automatically have the gaugino condensation
scale in the right ballpark, that is around $10^{12}$-$10^{13}$ GeV
\cite{Lebedev:2006tr}.  For 3 Wilson lines, we present  the statistics of the
hidden sector gauge groups in
Fig.~\ref{fig:ComparingHiddenSector3WLVs2-3WL}~(a).
There $N$ labels the ``size'' of the gauge groups  $\SU{N}$ and $\SO{2N}$.
The peak of this distribution is at $N=4$, which  leads  (in the absence of
hidden matter) to  gaugino condensation at an intermediate scale. If SUSY
breaking is due to gaugino condensation, the corresponding soft masses are in
the TeV range as favored by phenomenology.

\begin{center}
\begin{figure}[!h!]
\centerline{\includegraphics{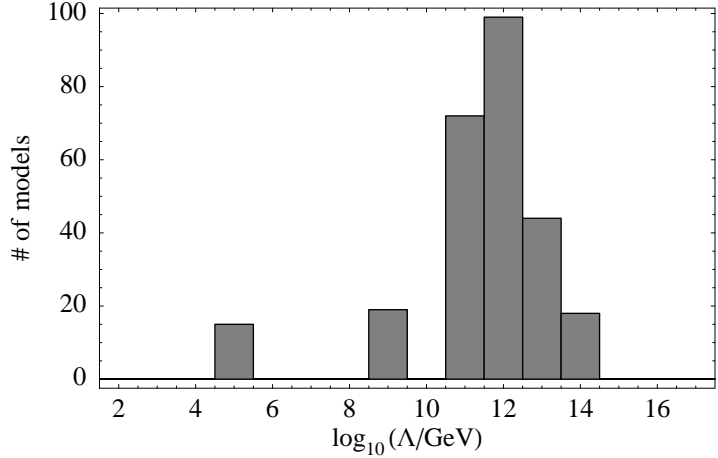}}
\caption{Number of MSSM candidates with 2 and 3 Wilson lines  vs.\ the
scale of gaugino condensation.}
\label{fig:scale}
\end{figure}
\end{center}

Combining both 2 and 3 WL models, we get again a distribution peaking at $N=4$
(Fig.~\ref{fig:ComparingHiddenSector3WLVs2-3WL}~(b)).  The corresponding gaugino
condensation scales are plotted in Fig.~\ref{fig:scale}. It is remarkable that
requiring the exact MSSM spectrum in the observable sector constrains  the
hidden sector such that gaugino condensation at an intermediate scale is
automatically preferred. This provides a top-down motivation for TeV scales in
particle physics.

\subsection{Example}

The model is defined by the gauge shift and Wilson lines
\begin{subequations}
\begin{eqnarray}
V^{\E6,1} & = &\left(\tfrac{1}{2},\,\tfrac{1}{3},\,\tfrac{1}{6},\,0,\,0,\,0,\,0,\,0\right)
               \left(0,\,0,\,0,\,0,\,0,\,0,\,0,\,0\right) \;,\\
W_{2} & = &\left(-\tfrac{1}{2},\,1,\,-\tfrac{1}{2},-1,\,0,\,0,-\tfrac{1}{2},\,-\tfrac{1}{2}\right)
           \left(-\tfrac{3}{4},\,-\tfrac{1}{4},\,-\tfrac{1}{4},\,-\tfrac{1}{4},\,-\tfrac{1}{4},\,\tfrac{1}{4},\,\tfrac{1}{4},\,\tfrac{1}{4},\,\right)\;,\\ 
W_{2}'& = &\left(\tfrac{1}{4},\,-\tfrac{1}{4},\,-\tfrac{1}{4},-\tfrac{7}{4},\,\tfrac{1}{4},\,-\tfrac{3}{4},\tfrac{1}{4},\,\tfrac{5}{4}\right)
           \left(0,\,-1,\,\tfrac{1}{2},\,\tfrac{1}{2},\,1,\,0,\,1,\,1\right)\;,\\ 
W_{3} & = &\left(-\tfrac{1}{6},\,\tfrac{1}{2},\,-\tfrac{1}{2},\tfrac{5}{6},\,-\tfrac{1}{6},\,-\tfrac{1}{6},-\tfrac{1}{6},\,-\tfrac{1}{6}\right)
           \left(\tfrac{2}{3},\,-1,\,0,\,0,\,0,\,-\tfrac{1}{3},\,0,\,\tfrac{2}{3}\right)\;.
\end{eqnarray}
\end{subequations}
It has an \E6 local GUT at the origin of the torus lattice.
The gauge group after compactification is
\begin{equation}
G_\mathrm{SM}\times[\SU3\times\SU5]\times\U1^6\;,
\end{equation}
where $G_\mathrm{SM}=\SU3_C\times\SU2_\mathrm{L}\times\U1_Y$ includes the standard \SU5 hypercharge generator
\begin{equation}
\mathsf{t}_Y~=~
\left(0,0,0,\tfrac{1}{3},\tfrac{1}{3},\tfrac{1}{3},-\tfrac{1}{2},-\tfrac{1}{2}\right) \, (0,0,0,0,0,0,0,0)\;.  
\end{equation}
The resulting massless spectrum is displayed in table~\ref{tab:spectrum}. One of
the SM families comes from the $\boldsymbol{27}$--plet of \E6 at the origin,
while the other two come  from various twisted and untwisted sectors. All three
generations are intrinsically different in this model.

\begin{table}[!h!]
\begin{center}
\begin{tabular}{|rlc|rlc|c|rlc|}
\cline{1-6}\cline{8-10}
  \#  &  Irrep                                          & Label & \# & Anti-irrep       & Label    &&  \#  & Irrep & Label \\
\cline{1-6}\cline{8-10}
  4 & $( {\bf 3}, {\bf 2};  {\bf 1},  {\bf 1})_{1/6}$   & $q_i$ &
  1 & $( {\bf 3}, {\bf 2};  {\bf 1},  {\bf 1})_{1/6}$   & $\bar q_i$ &&
 26 & $( {\bf 1}, {\bf 1};  {\bf 1},  {\bf 1})_{0}$     & $s^0_i$ \\
 14 & $( {\bf 1}, {\bf 2};  {\bf 1},  {\bf 1})_{-1/2}$  & $\ell_i$ &
  8 & $( {\bf 1}, {\bf 2}; {\bf 1}, {\bf 1})_{1/2}$     & $\bar\ell_i$ && 
 10 & $( {\bf 1}, {\bf 1};  {\bf\overline{3}}, {\bf 1})_{0}$ & $\bar h_i$ \\
  1 & $( {\bf 1}, {\bf 2};  {\bf\overline 3},  {\bf 1})_{-1/2}$  & $\ell'_i$ &
  2 & $( {\bf 1}, {\bf 2}; {\bf 3}, {\bf 1})_{1/2}$      & $\bar\ell'_i$ && 
  5 & $( {\bf 1}, {\bf 1};  {\bf 3}, {\bf 1})_{0}$       & $h_i$ \\
  4 & $( {\bf\overline{3}}, {\bf 1}; {\bf 1}, {\bf 1})_{-2/3}$  & $\bar u_i$ &
  1 & $( {\bf 3}, {\bf 1}; {\bf 1}, {\bf 1})_{2/3}$      & $u_i$ && 
  1 & $( {\bf 1}, {\bf 1};  {\bf 1}, {\bf 5})_{0}$       & $w_i$\\
  4 & $( {\bf 1}, {\bf 1};  {\bf 1}, {\bf 1})_{1}$       & $\bar e_i$ &
  1 & $( {\bf 1}, {\bf 1};  {\bf 1}, {\bf 1})_{-1}$      & $e_i$ && 
  1 & $( {\bf 1}, {\bf 1};  {\bf 1}, {\bf\overline{5}})_{0}$ & $\bar w_i$ \\
\cline{8-10}
 13 & $( {\bf \overline{3}}, {\bf 1};  {\bf 1}, {\bf 1})_{1/3}$ & $\bar d_i$ &
  7 & $( {\bf 3}, {\bf 1}; {\bf 1}, {\bf 1})_{-1/3}$     & $d_i$ &\multicolumn{3}{c}{$\phantom{I^{I^I}}$}\\
  1 & $( {\bf \overline{3}}, {\bf 1};  {\bf\overline{3}}, {\bf 1})_{1/3}$ & $\bar d'_i$ &
  2 & $( {\bf 3}, {\bf 1}; {\bf 3}, {\bf 1})_{-1/3}$     & $d'_i$ &\multicolumn{3}{c}{$\phantom{I^{I^I}}$}\\
\cline{1-6}
  2 & $( {\bf 3},  {\bf 1};  {\bf 1},  {\bf 1})_{1/6}$   & $v_i$ &  
  2 & $( {\bf\overline{3}}, {\bf 1}; {\bf 1}, {\bf 1})_{-1/6}$  & $\bar v_i$ & \multicolumn{3}{c}{$\phantom{I^{I^I}}$} \\
  2 & $( {\bf 1},  {\bf 1};  {\bf 3},  {\bf 1})_{1/2}$   & $\tilde s^+_i$ &
  2 & $( {\bf 1}, {\bf 1}; {\bf\overline{3}}, {\bf 1})_{-1/2}$ & $\tilde s^-_i$ & \multicolumn{3}{c}{$\phantom{I^{I^I}}$} \\
  8 & $( {\bf 1},  {\bf 1};  {\bf 1},  {\bf 1})_{1/2}$   & $s^+_i$ &  
  8 & $( {\bf 1},  {\bf 1};  {\bf 1},  {\bf 1})_{-1/2}$  & $s^-_i$ &  \multicolumn{3}{c}{$\phantom{I^{I^I}}$} \\
  4 & $( {\bf 1},  {\bf 2};  {\bf 1},  {\bf 1})_{0}$     & $m_i$ &  \multicolumn{3}{c|}{$\phantom{I^{I^I}}$} &  \multicolumn{3}{c}{$\phantom{I^{I^I}}$} \\
\cline{1-6}
\end{tabular}
\caption{Massless spectrum. Representations  with respect to
$[\SU3_C\times\SU2_\mathrm{L}]\times[\SU3\times\SU5]$ are given  in  bold face, the hypercharge is
indicated by the subscript.}
\label{tab:spectrum}
\end{center}
\end{table}

At this stage, the model has three generations of SM matter plus vector-like exotics. 
Once the SM singlets 
$s_i =\left\{s^0_i,\,h_i,\,\bar h_i\right\}$ develop nonzero VEVs, the gauge group breaks to
\begin{equation}
G_\mathrm{SM}\times G_\mathrm{hidden}
\end{equation}
with $G_\mathrm{hidden}=\SU5$.  Furthermore, we have verified that the mass
matrices of the vector-like exotics have maximal rank. Therefore, all the
exotics decouple from the low energy theory.  These mass matrices are given by 
\begin{equation*}
\mbox{\tiny $
\begin{array}{rclcrcl}
\mathcal{M}_{\ell\bar\ell}
& \!\!\!\!\!\!= &
\!\!\!\!\!\!\left(
\begin{array}{cccccccccc}
s^3 & s^1 & s^1 & s^5 & s^5 & s^5 & s^5 & s^1 & s^1 & s^2\\
s^1 & s^5 & s^5 & 0 & 0 & 0 & 0 & s^3 & s^3 & s^3\\
s^3 & s^4 & s^4 & s^4 & s^5 & s^5 & s^5 & s^1 & s^1 & s^2\\
s^3 & s^4 & s^4 & s^5 & s^4 & s^5 & s^5 & s^1 & s^1 & s^2\\
s^3 & s^4 & s^4 & s^5 & s^5 & s^4 & s^5 & s^1 & s^1 & s^2\\
s^3 & s^4 & s^4 & s^5 & s^5 & s^5 & s^4 & s^1 & s^1 & s^2\\
s^3 & s^4 & s^4 & s^4 & s^5 & s^5 & s^5 & s^1 & s^1 & s^2\\
s^3 & s^4 & s^4 & s^5 & s^4 & s^5 & s^5 & s^1 & s^1 & s^2\\
s^3 & s^4 & s^4 & s^5 & s^5 & s^4 & s^5 & s^1 & s^1 & s^2\\
s^3 & s^4 & s^4 & s^5 & s^5 & s^5 & s^4 & s^1 & s^1 & s^2\\
s^1 & s^3 & s^3 & s^6 & s^6 & s^6 & s^6 & s^4 & s^4 & s^3\\
s^4 & s^1 & s^1 & s^4 & s^6 & s^5 & s^5 & s^3 & s^3 & s^4\\
s^4 & s^1 & s^1 & s^6 & s^4 & s^5 & s^5 & s^3 & s^3 & s^4\\
s^4 & s^1 & s^1 & s^5 & s^5 & s^4 & s^6 & s^3 & s^3 & s^4\\
s^4 & s^1 & s^1 & s^5 & s^5 & s^6 & s^4 & s^3 & s^3 & s^4\\
\end{array}
\right)\;,
&
\mathcal{M}_{\bar d d}
& \!\!\!\!\!\!= &
\!\!\!\!\!\!\!\left(
\begin{array}{ccccccccc}
s^4 & s^4 & s^5 & s^5 & s^5 & s^5 & s^3 & s^3 & s^3\\
s^4 & s^4 & s^4 & s^5 & s^5 & s^5 & s^1 & s^1 & s^2\\
s^4 & s^4 & s^5 & s^4 & s^5 & s^5 & s^1 & s^1 & s^2\\
s^4 & s^4 & s^5 & s^5 & s^4 & s^5 & s^1 & s^1 & s^2\\
s^4 & s^4 & s^5 & s^5 & s^5 & s^4 & s^1 & s^1 & s^2\\
s^4 & s^4 & s^4 & s^5 & s^5 & s^5 & s^1 & s^1 & s^2\\
s^4 & s^4 & s^5 & s^4 & s^5 & s^5 & s^1 & s^1 & s^2\\
s^4 & s^4 & s^5 & s^5 & s^4 & s^5 & s^1 & s^1 & s^2\\
s^4 & s^4 & s^5 & s^5 & s^5 & s^4 & s^1 & s^1 & s^2\\
s^3 & s^3 & s^6 & s^6 & s^6 & s^6 & s^4 & s^4 & s^2\\
s^1 & s^1 & s^4 & s^6 & s^5 & s^5 & s^3 & s^3 & s^4\\
s^1 & s^1 & s^6 & s^4 & s^5 & s^5 & s^3 & s^3 & s^4\\
s^1 & s^1 & s^5 & s^5 & s^4 & s^6 & s^3 & s^3 & s^4\\
s^1 & s^1 & s^5 & s^5 & s^6 & s^4 & s^3 & s^3 & s^4\\
\end{array}
\right)\;,
\end{array}
$}
\end{equation*}
\begin{equation*}
\mbox{\footnotesize $
\begin{array}{rcll}
\mathcal{M}_{s^+s^-}
& \!\!\!\!\!\!= &
\!\!\!\!\!\!\left(
\begin{array}{cccccccccc}
s^4 & s^5 & s^4 & s^4 & s^5 & s^6 & s^5 & s^5 & s^4 & s^5\\
s^5 & s^4 & s^4 & s^4 & s^5 & s^4 & s^5 & s^5 & s^6 & s^5\\
s^5 & s^5 & s^6 & s^6 & s^1 & s^6 & s^5 & s^6 & s^6 & s^6\\
s^5 & s^5 & s^6 & s^6 & s^6 & s^4 & s^4 & s^6 & s^4 & s^4\\
s^5 & s^5 & s^6 & 0 & s^4 & s^6 & s^3 & s^4 & s^6 & s^4\\
s^5 & s^5 & s^6 & s^6 & s^6 & s^6 & s^6 & s^1 & s^6 & s^5\\
s^5 & s^5 & s^6 & s^6 & s^6 & s^4 & s^4 & s^6 & s^4 & s^4\\
s^5 & s^5 & 0 & s^6 & s^4 & s^6 & s^4 & s^4 & s^6 & s^3\\
s^4 & s^4 & s^4 & s^6 & 0 & 0 & s^6 & s^6 & 0 & s^6\\
s^4 & s^4 & s^6 & s^4 & s^6 & 0 & s^6 & 0 & 0 & s^6\\
\end{array}
\right)\,,
&
\begin{array}{rcl}
\mathcal{M}_{q\bar q}
& = &
\left(
\begin{array}{cccc}
s^3 & s^3 & s^3 & s^2
\end{array}
\right)\;, \\
\mathcal{M}_{u\bar u}
& = &
\left(
\begin{array}{cccc}
s^4 & s^3 & s^3 & s^3\\
\end{array}
\right)\;, \\
\mathcal{M}_{e\bar e}
& = &
\left(
\begin{array}{cccc}
s^4 & s^3 & s^3 & s^3\\
\end{array}
\right)\;, \\
\mathcal{M}_{v\bar v}
& = &
\left(
\begin{array}{cc}
s^5 & s^6\\
s^6 & s^5\\
\end{array}
\right)\;, \\
\mathcal{M}_{mm}
& = &
\left(
\begin{array}{cccc}
0 & s^6 & s^5 & s^4\\
s^6 & 0 & s^4 & s^5\\
s^5 & s^4 & 0 & s^6\\
s^4 & s^5 & s^6 & 0\\
\end{array}
\right)\;,
\end{array}
\end{array}
$}
\end{equation*}
where  $\ell$ collectively refers to $\ell_i$ and  $\ell'_i$, etc. 
At the same time, the hidden sector $\boldsymbol{5}$-plets acquire large  masses
and decouple.

We note that the model allows for an order one top Yukawa coupling. It results
from the couplings  of the type $U\,U\,U$ and $U\,T\,T$. Also, in this model,
one can define a non-anomalous $B-L$ symmetry which gives standard charges for
the SM matter. This feature is desired for proton stability, see
\cite{Buchmuller:2006ik,Lebedev:2006kn,Lebedev:2007hv,Buchmuller:2008uq}.

\section{Conclusions}

We have completed our search for MSSMs in the \Z6-II orbifold by including
models with 3 Wilson lines.  Out of a total of $10^7$,  we have identified 
almost 300 inequivalent models with the exact MSSM spectrum, gauge coupling
unification and a heavy top quark.  Models with  these features  originate
predominantly  from  \SO{10}, \E6 and \SU5   local GUTs.  Therefore, local GUTs
are instrumental in obtaining the right models. We also find that the scale of
gaugino condensation in the hidden sector is correlated with properties of the
observable sector such that soft masses in the TeV range are preferred. 

\section*{Acknowledgements}

This research was supported by the DFG cluster of excellence Origin and
Structure of the Universe, the European Union 6th framework program
MRTN-CT-2004-503069 "Quest for unification", MRTN-CT-2004-005104
"ForcesUniverse", MRTN-CT-2006-035863 "UniverseNet", SFB-Transregios 27
"Neutrinos and Beyond" and 33 "The Dark Universe" by Deutsche
Forschungsgemeinschaft (DFG). One of us (M.R.) would like to thank the Aspen
Center for Physics, where part of this work has been done, for hospitality and
support.

\bibliography{Orbifold}
\addcontentsline{toc}{section}{Bibliography}
\bibliographystyle{NewArXiv}

\end{document}